\documentclass[prb,twocolumn,showpacs,preprintnumbers,amsmath,amssymb]{revtex4}

\usepackage{graphicx}
\usepackage{dcolumn}
\usepackage{bm}

\begin{document}

\preprint{APS/123-QED}

\title{Diffusion limited cluster aggregation with irreversible flexible bonds}
\author{Sujin Babu}
\author{Jean Christophe Gimel}
\email{Jean-Christophe.Gimel@univ-lemans.fr}
\author{Taco Nicolai}%
\email{taco.nicolai@univ-lemans.fr}
\affiliation{Polym\`eres Collo\"{\i}des Interfaces, CNRS UMR6120,
Universit\'e du Maine, F-72085 Le Mans cedex 9, France}

\date{\today}
\begin{abstract}
Irreversible diffusion limited cluster aggregation (DLCA) of hard spheres was simulated using Brownian cluster dynamics. Bound spheres were allowed to move freely within a specified range, but no bond breaking was allowed. The structure and size distribution of the clusters was investigated before gelation. The pair correlation function and the static structure factor of the gels were determined as a function of the volume fraction and time. Bond flexibility led to local densification of the clusters and the gels, with a certain degree of order. At low volume fractions densification of the clusters occurred during their growth, but at higher volume fractions it occurred mainly after gelation. At very low volume fractions, the large scale structure (fractal dimension), size distribution and growth kinetics of the clusters was found to be close to that known for DLCA with rigid bonds. Restructuring of the gels continued for long times, indicating that aging processes in systems with strong attraction do not necessarily involve bond breaking. The mean square displacement of particles in the gels was determined. It is shown to be highly heterogeneous and to increase with decreasing volume fraction.
\end{abstract}

\maketitle
\section{Introduction}
Strong attraction between particles in solution leads to aggregation. The kinetics of this aggregation process depends on the probability that a bond is formed when two particles collide. Two limiting cases are diffusion limited cluster aggregation (DLCA) if bonds are formed at each collision and reaction limited cluster aggregation (RLCA) if the probability to form a bond is very small. Irreversible aggregation has been studied in detail both experimentally \cite{954,734,860,796,747,748} and using computer simulations \cite{193,194,943,171,340,454,802,803,seager,759,767}. The structure of clusters formed by random aggregation is self similar and characterized by a fractal dimension ($d_f$), which relates the radius of gyration ($R_g$) to the aggregation number ($m$): $m\propto R_{g}^{d_f}$. The number of clusters with aggregation number $m$ can be described by a power law: $N(m)\propto m^{-\tau}$. As long as the clusters are on average far apart (flocculation) one finds $d_f=1.8$ and $\tau=0$ for DLCA and $d_f=2.1$ and  $\tau=1.5$ for RLCA \cite{175}.

With time the clusters grow and the cumulated volume occupied by the clusters ($V_{cum}=\sum N(m) \cdot 4\pi R_{g}^{3}/3$) increases so that the average free space between the aggregates decreases. When $V_{cum}$ approaches the volume of the system the aggregates start to interpenetrate. The aggregation process of highly interpenetrated clusters can be described by the percolation model and leads to gelation. For percolating clusters $d_f=2.5$ and $\tau=2.2$ \cite{2}. The cross-over between flocculation and percolation occurs at a characteristic aggregation number ($m_c$) and radius of gyration ($R_c$), that decrease with increasing particle volume fraction ($\phi$) \cite{943,171,802}.

  Computer simulations of irreversible DLCA and RLCA have been done so far for hard spheres that form rigid bonds at contact \cite{802,803,767,530}. In this case the aggregated particles are on average bound to two other particles, because ternary collisions are not possible. However, in reality the bonds may be flexible, i.e. they may freely rotate. One example is the much studied aggregation of spheres in the presence of other smaller particles through a depletion interaction \cite{779,1044,1061,1018,1030,manely}. Another example is the aggregation of emulsion droplets with a slippery layer \cite{954}. The latter experiment has motivated computer simulations of irreversible diffusion limited aggregation (DLA) with flexible bonds \cite{seager}. The difference between DLA and DLCA is that during the former individual particles are allowed to diffuse until they collide with a single cluster \cite{175}, while during the latter all particles in the systems diffuse and collide to form many clusters. DLA leads to a self similar cluster with $d_f=2.5$. In ref. \cite{seager}  the DLA simulation was modified to include free diffusion of the particle on the surface of the particle to which it is bound and was called slippery DLA. Real random aggregation processes are, of course, better described by DLCA.
  
Here we report on a simulation study of DLCA with finite interaction range in which the relative motion of bound particles is unhindered as long they remain within each others range. If the interaction range is very small compared to the radius of the particles this method could be called slippery DLCA, but here we will use the expression flexible DLCA for all interaction ranges.  Flexible DLCA (or RLCA) should represent realistically the experimental systems mentioned above in the limiting case that the attraction energy is much larger than the thermal energy. The results of flexible DLCA will be compared to those obtained by DLCA with rigid bonds. 

\section{Simulation method}
The simulation method used here is called Brownian Cluster Dynamics (BCD). A detailed description of the method and a comparison with molecular dynamics were reported elsewhere \cite{babu}. Briefly, clusters are formed by connecting spheres within each others interaction range with probability $P$. Particles are chosen randomly and moved a step size $s$ in a random direction unless it leads to overlap or breaks a bond. The centre of mass displacement of the clusters is calculated and the clusters are moved cooperatively in the same direction so that the total displacement is inversely proportional to their radius, unless it leads to overlap. BCD is equivalent to molecular dynamics if the cooperative cluster displacement is omitted as long as $s$ is sufficiently small. Systems with rigid bonds are simulated by performing only the cooperative cluster movements and not the individual particle displacements within the clusters. The equilibrium state obtained by molecular dynamics and BCD is the same, but the dynamics depend strongly on whether the bonds are rigid or flexible.

Irreversible DLCA is simulated by setting $P=1$. The simulation is started with $N_{tot}$ randomly distributed spheres with unit diameter in a box of size $L$ so that $\phi=N_{tot}/L^3(\pi/6)$. The unit of time is set equal to the time needed for an isolated sphere to diffuse a distance equal to its diameter. E.g. for spheres with diameter $1\mu m$ in water at $20^{\circ}C$ the time unit is $0.4$ seconds. The box size was varied up to $L=100$ and all the results shown here were not influenced by finite size effects unless specified.  

\section{Results}
In the following we show mainly results obtained with the interaction range fixed at $\epsilon=0.1$, but we will briefly discuss the effect of varying the range.       

\subsection{Kinetics}
It is well known that in the flocculation regime, i.e. $V_{cum}\ll L^3$ the cluster growth during DLCA can be described by the kinetic equations introduced by Smolechowski \cite{191,192,214}: 
\begin{equation}
\frac{{\rm d}N(m)}{{\rm d}t}=\frac{1}{2}\sum_{i+j}K(i,j)N(i)N(j)-\sum_{j}K(m,j)N(m)N(j)
\label{e.1}
\end{equation}
Where $K(i,j)$ is the so-called kernel that expresses the rate constant at which a cluster with aggregation number $i$ collides with a cluster with aggregation number $j$: 
\begin{equation}
K(i,j)=4\pi \left({R_{col,i}+R_{col,j}}\right)\left({D_i+D_j}\right)
\label{e.2}
\end{equation}
with $D$ the diffusion coefficient of the clusters and $R_{col}$ their collision radius. $D$ is inversely proportional to the hydrodynamic radius ($R_h$) of the clusters. For large clusters both $R_{col}$ and $R_h$ are proportional to $R_g$ \cite{776}. In fact, the growth kinetics of DLCA can be well described by eq(\ref{e.1}) using a constant kernel equal to $K(1,1)$. It follows that the time dependence of the weight ($m_w$) averaged aggregation number can be written as:
\begin{equation}
m_w=1+K(1,1)t\phi
\label{e.3}
\end{equation}

In our simulations the particle diameter is unity and the diffusion coefficient of the individual spheres is $1/6$ so that $K(1,1)=4\pi /3$. The ratio of the weight and the number ($m_n$) averaged aggregation numbers becomes two for large $m_w$. These results have been confirmed using BCD with rigid bonds, but only if $\phi$ was very low. In \cite{802} the interaction range was zero. Increasing the interaction range increases the collision radius and therefore $K(1,1)$, but the effect is small for $\epsilon=0.1$. For $\phi >0.01$ it is difficult to observe the flocculation regime because the crossover to the percolation regime occurs already when $m_w$ is still relatively small. The growth of $m_w$ accelerates when $V_{cum}$ approaches $L^3$ and $m_w$ diverges at the gel time ($t_g$). This change of the growth kinetics between flocculation at short times and percolation close to $t_g$ has also been observed experimentally \cite{945}.

\begin{figure}
\resizebox{0.45\textwidth}{!}{\includegraphics{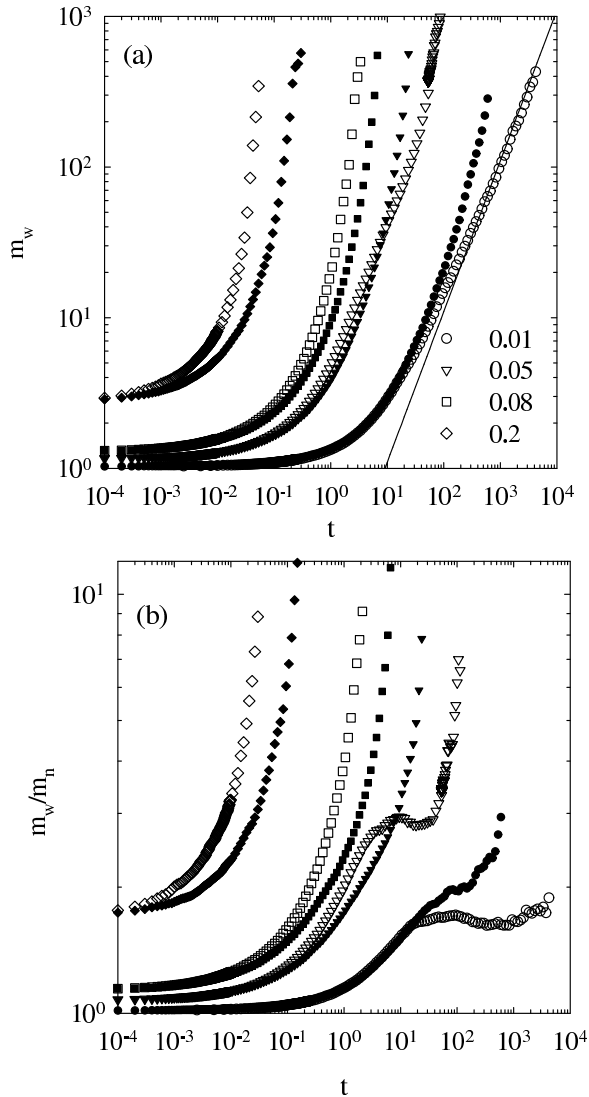}}
\caption{(a) $m_w$ is plotted as a function of time $t$ obtained from rigid DLCA (filled symbols) and flexible DLCA (open symbols) for $\epsilon=0.1$ at different $\phi$ as indicated in the figure. The solid line has a slope of $1$. (b) The polydispersity index is plotted as a funtion of $t$ for different $\phi$.}
\label{f.1}
\end{figure}

Figure(\ref{f.1}a) compares the increase of $m_w$ as a function of time for DLCA with rigid and with flexible bonds at different volume fractions, while fig.(\ref{f.1}b) shows the evolution of the polydispersity index ($m_w/m_n$). The size of the clusters at the start of the simulation depends on the interaction range and the volume fraction, because more particles are within each others range if $\epsilon$ or $\phi$ are larger. For large $\epsilon$ and $\phi$ the system percolates immediately, i.e. when the spheres are still randomly distributed, as was discussed in detail in \cite{750}. The percolation threshold decreases with increasing $\phi$ and is $0.27$ for $\epsilon=0.1$.

At $\phi=0.01$ the increase of $m_w$ with time was linear and $m_w/m_n$ was close to two as expected for the flocculation regime. The results were similar for rigid and flexible DLCA, but rigid DLCA showed the influence of a transition to the percolation regime sooner. The gel point was not reached for this volume fraction during the simulation which took several weeks of computer time. For $\phi >0.08$ the flocculation regime was not observed and the growth of both the size and the polydispersity of the clusters accelerated until they diverged at $t_g$. The growth of the clusters and gelation was somewhat faster for flexible DLCA. At $\phi=0.05$ the increase of $m_w$ was initially slightly faster during flexible DLCA, but then the growth became almost linear over a short period of time while the polydispersity index remained constant. After this period the increase of $m_w$ and $m_w/m_n$ accelerated until the gel was formed. During rigid DLCA the slowing down of the aggregation was not observed so that gelation at this volume fraction was faster than for flexible DLCA. The growth rate at different volume fractions can be understood by considering the structure of the clusters as will be discussed below.

\begin{figure}
\resizebox{0.45\textwidth}{!}{\includegraphics{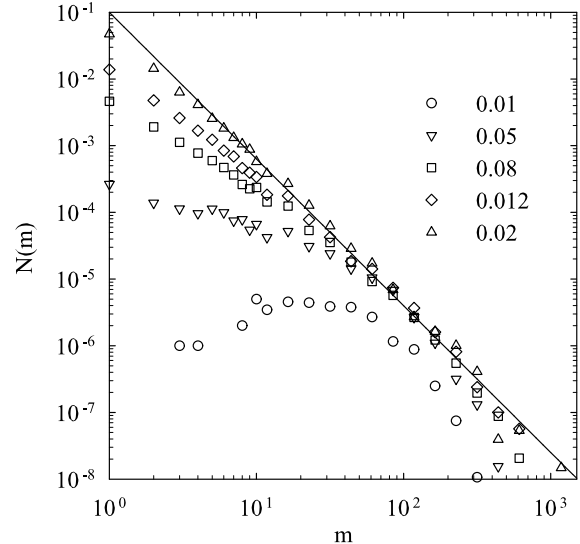}}
\caption{Cluster size distribution is plotted as a function of the aggregation number $m$ obtained from flexible DLCA for $\epsilon=0.1$ at different $\phi$ as indicated in the figure. The solid line has a slope of $-2.2$.}
\label{f.2}
\end{figure}

Figure(\ref{f.2}) shows $N(m)$ for $m_w=100$ obtained at different volume fractions. At the lowest concentration we observed a bell shape as is also observed for rigid DLCA in the flocculation regime \cite{802,852}. At higher volume fractions $N(m)$ followed approximately the power law decay expected for the percolation regime: $N(m)\propto m^{-2.2}$. We note that the cut-off function of the size distribution of percolating clusters has a Gaussian shape \cite{450,454}. More extensive simulations on even larger boxes will be needed to test whether $N(m)$ obtained from flexible DLCA has exactly the same functional form as for rigid DLCA or static percolation.   

\begin{figure}[h]
\resizebox{0.45\textwidth}{!}{\includegraphics{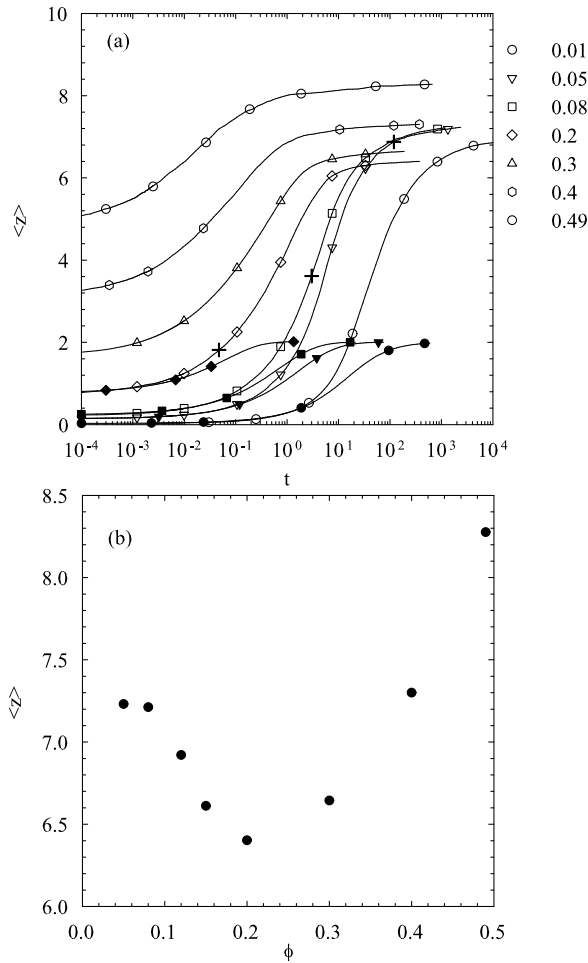}}
\caption{(a) The average number of neighbors obtained from rigid DLCA (filled symbols) and flexible DLCA (open symbols) for $\epsilon=0.1$ is plotted as a function of time for different $\phi$ as indicated in the figure. Crosses indicate the gel time for flexible DLCA. (b) $<z>$ at the end of the simulation for flexible DLCA with $\epsilon=0.1$ is plotted as a function of $\phi$.}
\label{f.3}
\end{figure}

Aggregation leads to an increase of the number of bonds per particle ($z$). In fig.(\ref{f.3}a) the average number of bonds per particle ($<z>$) is plotted as a function of time. As mentioned above, the starting value of $<z>$ is larger for larger $\phi$ because more randomly distributed spheres are within each others interaction range. At low volume fractions, $<z>$ increased during rigid DLCA until it stagnated at a value close to $2$, while at large $\phi$ the increase of $<z>$ was small because almost all the particles were already part of the final structure at the start of the aggregation. During flexible DLCA $<z>$ increased sharply at first, but when it had reached a value around $7$ further growth of $<z>$ became very slow. Contrary to the case of rigid DLCA, $<z>$ continued to increase over the whole duration of the simulation even after all spheres were attached to the network indicating that restructuring of the system occurred. This process was very slow and may be called ageing. For $\phi >0.08$ most of the restructuring happened after the gel point and continued even when the sol fraction had become very small. 

In fig.(\ref{f.3}b) are plotted the values of $<z>$ at the end of the simulation as a function of the volume fraction. $<z>$ has a minimum at $\phi \approx 0.2$, the origin of which will be discussed below. The distribution of $z$ at long times only weakly depended on the volume fraction for $ \phi<0.3$. It was approximately Gaussian with a half width of $4$. It started at $z=3$ and peaked close to $<z>$. Very few particles had $z=12$ showing that no crystallisation had occurred. At higher volume fractions the distribution shifted to larger $z$. 

\subsection{Cluster structure} 
During rigid DLCA the cluster configuration is determined by random collisions and the clusters cannot rearrange once they are formed. Therefore rigid DLCA yields in the flocculation regime clusters with the same average radius of gyration for a given aggregation number independent of time. This is not the case for flexible DLCA where the particles in the clusters rearrange until a maximum of bonds are formed without breaking any existing bond. In this way the density of the clusters increases with time. At very low volume fractions the growth rate was slower than the restructuring time so that at each moment the cluster configuration had reached the steady state, while at high volume fractions the gel was formed quicker than the time needed to restructure. Therefore the value of $<z>$ at the gel point was close to the final value for $\phi <0.05$, but much lower at large $\phi$, see fig.(\ref{f.3}a). 

\begin{figure}[h]
\resizebox{0.45\textwidth}{!}{\includegraphics{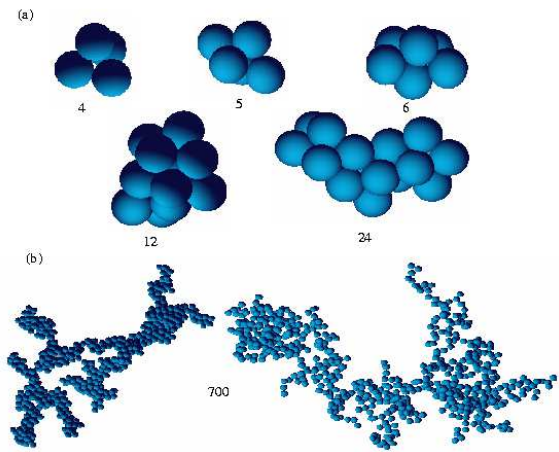}}
\caption{(a)Images of clusters with different aggregation numbers formed by flexible DLCA with $\epsilon=0.1$. (b) Comparison of a cluster formed by flexible DLCA (left) and by rigid DLCA (right) with aggregation number 700.}
\label{f.4}
\end{figure}

Figure(\ref{f.4}a) shows images of clusters formed during flexible DLCA at very low volume fractions. The configuration of clusters up to $m=6$ was unique contrary to clusters formed by rigid DLCA. For $m=7$ one configuration was observed in $90\%$ of the cases, but one other configuration was also possible. With increasing aggregation number more different configurations were found. Large clusters were generally elongated strands that branched when $m$ exceeded about $20$. The randomly branched structures resembled clusters formed by rigid DLCA, except that the strands were much thicker than the single particle diameter, see fig.(\ref{f.4}b). The basic unit of all larger clusters formed by flexible DLCA was a tetrahedron and for $m\geq 4$ particles with less than $3$ neighbours were not observed. The fact that the tetrahedron is the basic structural unit does not imply however that tetrahedra were formed first and subsequently aggregated to form larger clusters as was suggested in \cite{954}. At all stages of the aggregation we observed all aggregation numbers with no preference for multiples of $4$.

\begin{figure}[h]
\resizebox{0.45\textwidth}{!}{\includegraphics{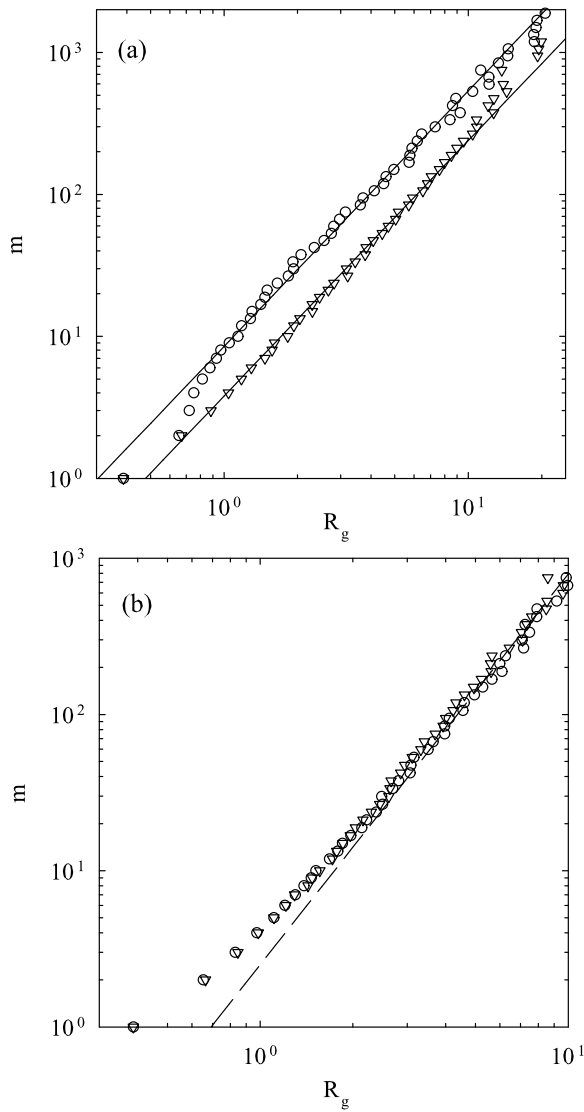}}
\caption{The aggregation number is plotted as function of $R_g$ for both rigid DLCA (triangles) and flexible DLCA (circles) at (a) $\phi=0.01$ and (b) $\phi=0.2$ with $\epsilon=0.1$. The solid lines and the dashed line have slopes of $-1.8$ and $-2.5$ respectively.}
\label{f.5}
\end{figure}

As mentioned in the introduction, the large scale structure of clusters formed by rigid DLCA in the flocculation regime is self similar with $d_f=1.8$. In fig.(\ref{f.5}a) a comparison is shown of the dependence of $m$ on $R_g$ between clusters formed by flexible DLCA and by rigid DLCA at $\phi=0.01$. In both cases $m=a \cdot R_{g}^{1.8}$, but the prefactor was larger for flexible DLCA ($a=8.5$) than for rigid DLCA ($a=3.8$) because the local structure of clusters is denser. Figure(\ref{f.5}b) shows the dependence of $m$ on $R_g$ for clusters formed at $\phi=0.2$ close to the gel point. The structure of the clusters was almost the same for flexible and rigid DLCA because not much restructuring had yet occurred before the gel was formed. For larger clusters $m\propto R_{g}^{2.5}$ as predicted by the percolation theory. At intermediate volume fractions the density of the clusters increased with time for flexible DLCA and the transition between flocculation and percolation shifted to larger values of $m$ with decreasing $\phi$.

The structure of the clusters at higher volume fractions was the same for rigid and flexible DLCA, yet the growth rate was faster for flexible DLCA. This can be explained by the fact that neighbouring fractal clusters have a larger probability to bind when they are flexible than when they are rigid. It is likely that in the case of rigid clusters rotation would also accelerate the growth during the percolation regime. At low volume fractions both the collision radius and the hydrodynamic radius of clusters formed during flexible DLCA were smaller for a given aggregation number. Apparently, both effects compensated to give similar growth kinetics for flexible and rigid DLCA. However, the percolation regime was reached at larger $m$ during flexible DLCA than during rigid DLCA. Densification of the clusters during flexible DLCA explains why the growth rate at $\phi=0.05$ was initially faster and at later times slower than for rigid DLCA. Initially, the growth was faster due to flexibility, but later densification rendered the clusters smaller so that the transition to the percolation regime was delayed.

\subsection{Gel structure}
The gel structure can be characterized by the pair correlation function ($g(r)$) or its Fourier transform the static structure factor ($S(q)$). $g(r)$ represents the average number concentration of particles at a distance $r$ from any given particle and reaches the number concentration of the system $(C=\phi 6/\pi)$ at large $r$. An extensive study of gels formed by rigid DLCA was reported in \cite{802,803}. It was shown that pair correlation functions of gels at low volume fractions had distinct features at small $r$ followed by a power law decay ($g(r)\propto r^{(d_f-3)}$ with $d_f=1.8$) for $r>3$. $g(r)$ had a weak minimum at a characteristic value $r_{min}$ close to the correlation length of the concentration fluctuations before it reached $C$. The correlation length decreased with increasing volume fraction, and became of the order of a few particle diameters for $\phi >0.05$. As a consequence, the fractal structure did not exist at higher volume fractions. 

$g(r)$ showed a delta peak at $r=1$ representing the contribution of the two nearest neighbours and increased continuously to $0.18$ at $r=2$ starting from a low value close to $r=1$. At low volume fractions the main contribution to $g(r)$ in this range came from the bound next-nearest neighbours. Their contribution stopped at $r=2$ which caused a discontinuous drop of $g(r)$. A small inflection of $g(r)$ was observed at $r=3$ marking the influence of the second shell and at larger $r$ the power law decay started. The local structure of the gels was identical for small volume fractions ($\phi <0.05$) but changed at higher volume fractions because positional correlations between randomly distributed spheres at $t=0$ were no longer negligible.

\begin{figure}[h]
\resizebox{0.45\textwidth}{!}{\includegraphics{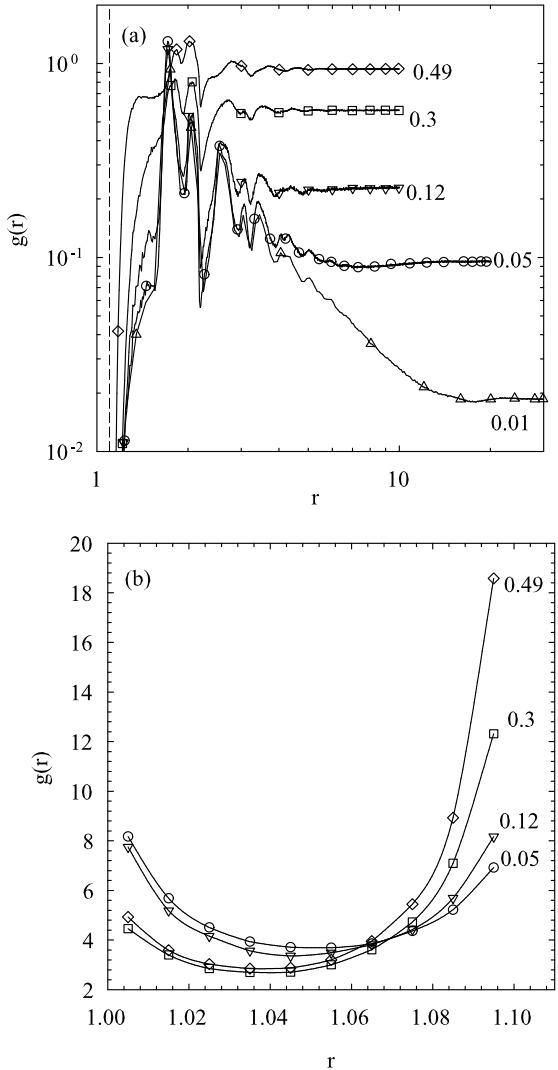}}
\caption{(a) The pair correlation function is plotted for different $\phi$ for $\epsilon=0.1$ as indicated in the figure. The dashed line indicate the interaction range $1.1$. (b) Zoom of $g(r)$ between $1<r<1+\epsilon$.}
\label{f.6}
\end{figure}

Figure.(\ref{f.6}a) shows pair correlation functions of gels obtained by flexible DLCA at different volume fractions. The values of $g(r)$ within the interaction range were much larger than for $r>1+\epsilon$ and are shown separately in fig.(\ref{f.6}b). There is a discontinuity in $g(r)$ at $r=1+ \epsilon$, i.e. the maximum distance between a pair of bonded particles. This discontinuity is a consequence of the square well interaction and is not seen for a continuous potential such as the Lennart-Jones potential. For $r>1+\epsilon$, peaks can be seen indicating a high degree of local order that was absent for rigid DLCA. The maxima at $r=1.7-1.8$ and $r=2-2.2$ are characteristic for the tetrahedral structure with a bond length between $1$ and $1.1$. If this structure is extended in a regular linear fashion it leads to the so-called Bernal spiral. Small peaks at larger distances indicate that the order persists to some extent but for $r>5$ the structure became self similar.

At even larger distances a very weak minimum was observed at small volume fractions similar to that found for rigid DLCA, but at a given volume fraction it was situated at larger $r$ values for flexible DLCA. At a given volume fraction, the correlation length of the gel is larger for flexible DLCA because the structure is locally denser. For the same reason the radius $R_c$ at the crossover between flocculation and percolation was larger. It was shown in \cite{803} for rigid DLCA, that the correlation length and $R_c$ are close. 

Bonded nearest neighbours can be situated at any distance within the interaction range, i.e. between $1$ and $1.1$ in the present case, but fig(\ref{f.6}b) shows that the distance was not uniformly distributed. In fact there was a preference for distances close to $1$ or $1+ \epsilon$, which means that some bonds were compressed and others stretched. The local structure was independent of the volume fraction for $\phi<0.1$, because it is mainly determined by the aggregation process of particles that were initially outside each others range. At higher volume fractions a significant number of particles were in contact before the aggregation process started. The restructuring was thus more constrained and leading to less local order. The split between the peaks at $r=1.7$ and $2$ became less distinct and resembled more closely the split peak observed for super cooled liquids.
 
\begin{figure}[h]
\resizebox{0.45\textwidth}{!}{\includegraphics{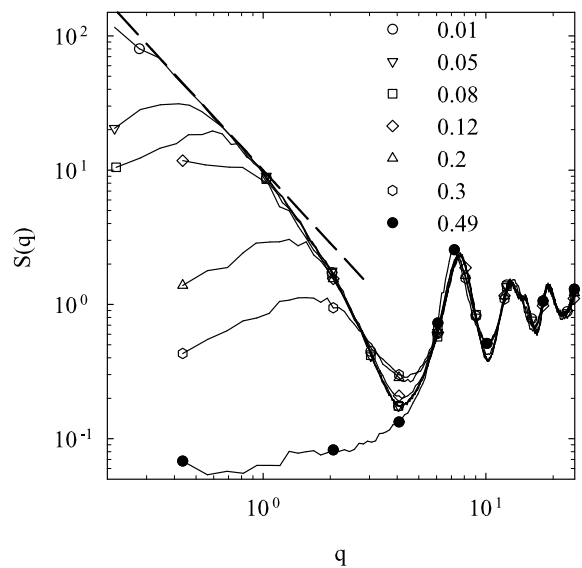}}
\caption{The structure factor is plotted for $\epsilon$ at different $\phi$ as indicated in the figure. The solid lines are guide to the eye and dashed line have a slope of $1.8$. For clarity we have kept only few symbols.}
\label{f.7}
\end{figure}

The corresponding structure factors are shown in fig.(\ref{f.7}) for different volume fractions. At small $q$ a maximum was found at a value ($q_{max}$) that increased with increasing concentration. For rigid DLCA it was shown that the position of maximum is inversely proportional to that of the minimum of the pair correlation function $q_{max}\propto 3/r_{min}$ \cite{803}. Over a narrow $q$-range $S(q)$ decreased with increasing $q$ following a power law, which is expected for self similar structures: $S(q)\propto q^{-d_f}$. The data are compatible with $d_f=1.8$, see the solid line in fig.(\ref{f.7}), but for a precise determination of the fractal dimension even smaller volume fractions need to be investigated. At large $q$-values $S(q)$ oscillated around unity.  

\begin{figure}[h]
\resizebox{0.45\textwidth}{!}{\includegraphics{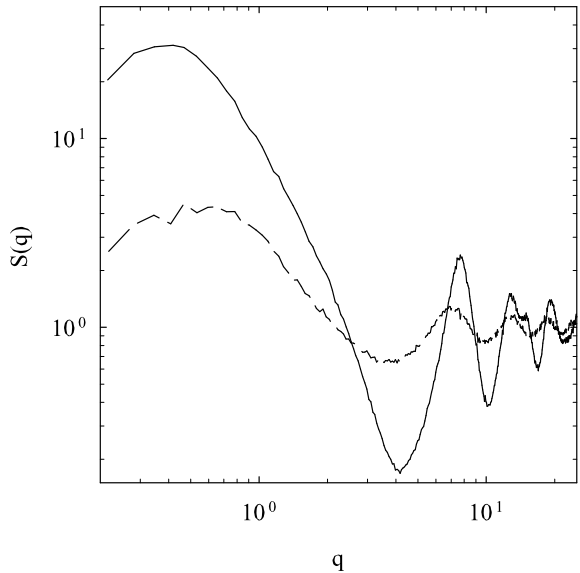}}
\caption{The structure factor obtained from flexible DLCA (solid lines) and rigid DLCA (dashed line) for $\phi=0.05$ and $\epsilon=0.1$.}
\label{f.8}
\end{figure}

Figure(\ref{f.8}) compares the structure factors of stable gels formed by flexible and rigid DLCA at $\phi=0.05$. The maximum has a much larger amplitude and is situated at smaller $q$-values for gels formed by flexible DLCA and the oscillations at large $q$ are stronger.  The reason is that the local density of the gels is higher and therefore the correlation length is larger. At this volume fraction a large scale fractal structure can be observed if the gels are produced by flexible DLCA, but not if they are produced by rigid DLCA.  

\subsection{Structural evolution} 
The evolution of $g(r)$ and $S(q)$ as a function of time during rigid DLCA was shown in \cite{803}. With time the maximum of $S(q)$ shifted to lower $q$ and increased in amplitude, while the minimum of $g(r)$ shifted to larger $r$, because the correlation length increased. During rigid DLCA the structure is fixed as soon as the bonds are formed, but this is not the case for flexible DLCA. We illustrate this point by showing the evolution of the structure at $\phi=0.08$ during flexible DLCA. At this volume fraction the local structure is not yet influenced by correlation between the randomly distributed spheres. 

\begin{figure}[h]
\resizebox{0.45\textwidth}{!}{\includegraphics{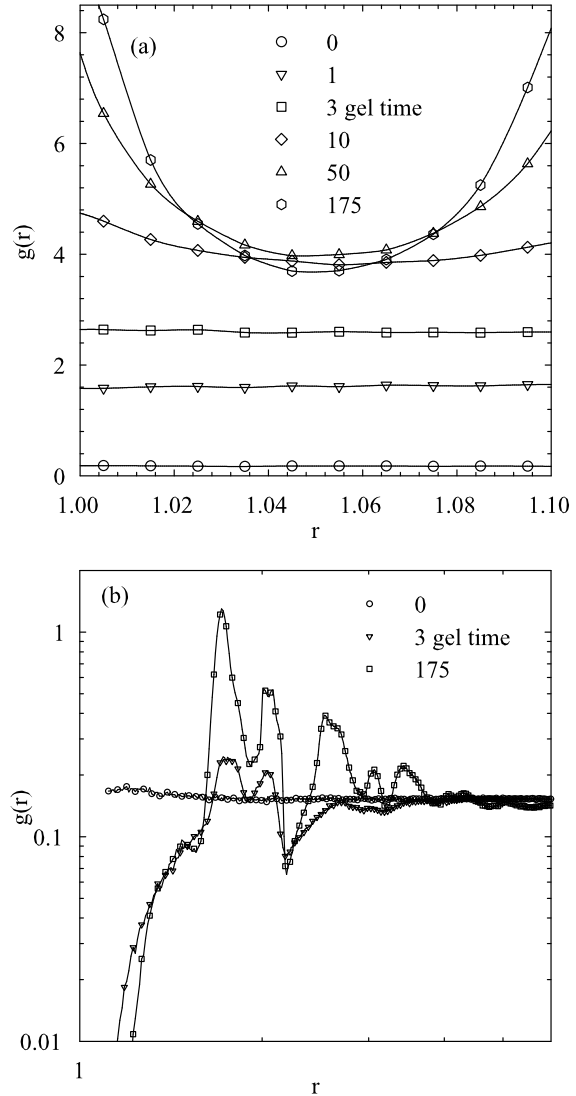}}
\caption{(a) $g(r)$ is plotted as a  funtion of distance for $\phi=0.08$ and $\epsilon=0.1$ between $1<r<1+\epsilon$ at different times as indicated in the figure. (b) $g(r)$ is plotted for $r>1+\epsilon$ for the same system at different times as indicated in the figure.}
\label{f.9}
\end{figure}

$g(r)$ is plotted in fig(\ref{f.9}) at different times. At the start of the aggregation, $g(r)$ was constant within the bond range, i.e. $1<r<1.1$, and equal to $C$. With time the number of bonds increased, but $g(r)$ remained constant in the range even when the gel was formed. At longer times, however, $g(r)$ became larger at $r=1$ and $r=1.1$, implying that in order to maximize the number of bonds under constraint, many bonds needed to be either stretched or compressed during restructuring of the gel. The peaks at $r=1.7$ and $r=2$ appeared already before the gel point indicating that some tetrahedral structure had been formed without bond stretching or compression. The characteristic minimum of $g(r)$ at $r_{min}$ developed slowly and moved to longer distances. 

\begin{figure}[h]
\resizebox{0.45\textwidth}{!}{\includegraphics{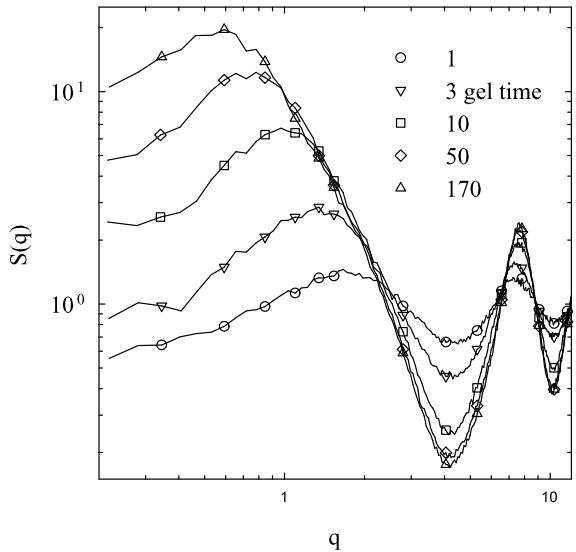}}
\caption{The $S(q)$ is plotted for $\phi=0.08$ and $\epsilon=0.1$ for different times as indicated in the figure.}
\label{f.10}
\end{figure}

Figure(\ref{f.10}) shows the corresponding evolution of the structure factor. The shift of the maximum to lower $q$ and the increase of its amplitude are caused by an increase of the correlation length. The amplitude of the oscillations at high $q$-values increased due to local densification.   
At low volume fractions the pair correlation function and the structure factor changed very little after the gel point because the aggregation was sufficiently slow so that the clusters had time to restructure before they percolated. At high volume fractions gels were formed very rapidly and the structure at the gel point resembled that of gels formed by rigid DLCA. In this case, the distinguishing features of the flexible DLCA gels developed after the gel point. At long times the structural changes became increasingly slow, but they did not stop completely during the simulation time. 

\subsection{Dynamics }
For rigid DLCA the mean squared displacement (MSD) of the spheres in the gels is, of course, zero. However, in flexible DLCA gels the spheres have a significant mobility, which can be characterised by measuring the MSD as a function of time. It is important to distinguish displacements due to restructuring from displacements due to flexibility. The displacement due to restructuring can be probed by measuring the average total MSD from the start of the aggregation process. The total MSD as a function of time increases rapidly at first followed by a very weak increase when all particles are part of the gel. 

\begin{figure}[h]
\resizebox{0.45\textwidth}{!}{\includegraphics{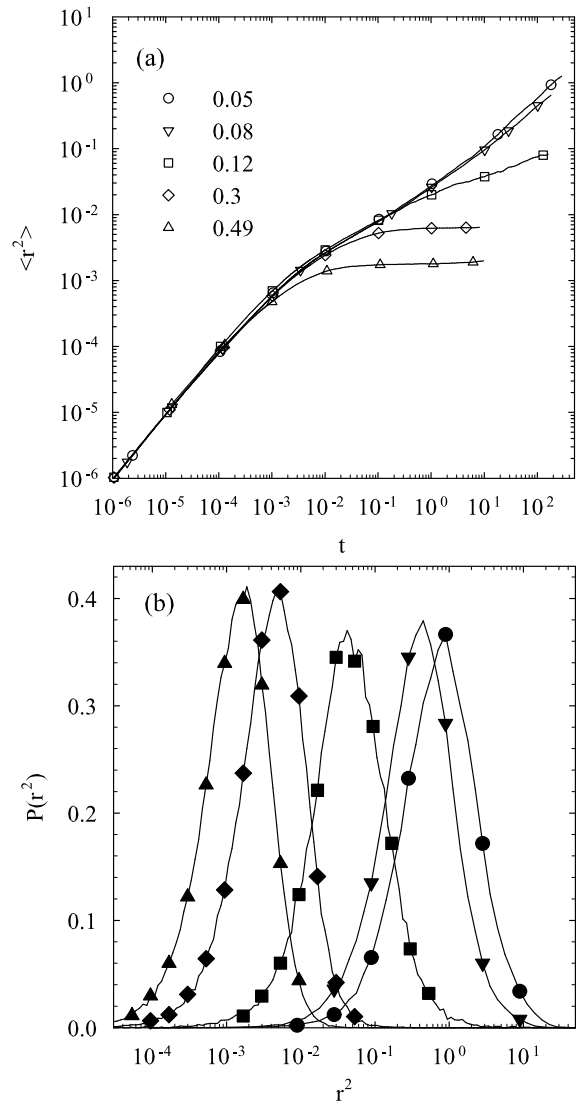}}
\caption{ (a) The MSD of spheres in aged gel is plotted as a funtion of time for different $\phi$ as indicated in the figure. (b) The corresponding distribution of $r^2$ at the longest time.}
\label{f.11}
\end{figure}

One can characterise the flexibility of the gels by measuring the MSD over a period of time during which restructuring is negligible, i.e. for gels that have aged for a long time. Alternatively, one can stop further bond formation and thus further restructuring. This is, of course, easier to do in simulations than in experiments on real systems. Figure(\ref{f.11}a) shows the MSD of spheres in gels at the latest simulation time, when all particles were part of the percolating network. Initially, free diffusion was observed over very short distances. Then the average MSD slowed down and finally stagnated at a value $<r^2>=\delta^2$. $\delta^2$ increased with decreasing concentration, but we could not properly observe $\delta^2$ at low volume fractions because the times scales involved were too large. 

	If the displacement of the particles is caused by Brownian motion, the distribution of $r^2$ (or the self part of the van Hove correlation function $P(r^2)$) can be described by a Gaussian function. Figure(\ref{f.11}b) shows $P(r^2)$ when the average was close to $\delta^2$. It is clear that the displacement was highly heterogeneous as might be expected from the heterogeneous structure of the gels. The non-Gaussian character can be expressed by the parameter $\alpha=<r^4>/(<r^2>)^2-5/3$ which is zero for a Gaussian function. We found $\alpha \approx 1$ for all volume fractions. 
	
The self part of the intermediate scattering function was found to fully decay at scattering wave vectors $q\gg 1/\delta$, while it showed a plateau at long delay times for smaller $q$ at a value that increased with decreasing $q$. This is, of course, a direct consequence of the restricted MSD of the particles, see below. 

\subsection{Effect of the interaction range. }
For rigid DLCA the effect of increasing the interaction range is twofold. In the first place, as mentioned above, the collision radius of the spheres is increased by a factor $1+\epsilon$. As a consequence the clusters are larger and the aggregation is faster. In the second place, the concentration of bonds between the randomly distributed spheres at the start of the aggregation is larger. The latter effect becomes important when the average distance between nearest neighbours ($\Delta $) is smaller than $1+\epsilon$  ($\Delta <1.1$ for $\phi>0.2$). The percolation threshold of randomly distributed spheres decreases with increasing $\epsilon$ was discussed in \cite{750}. However, as long as $\Delta$ is much larger than $1+\epsilon$, i.e. at small volume fractions, the effect of the interaction range on the structure of the clusters and the gels is small and the fractal dimension remains the same.
 
\begin{figure}[h]
\resizebox{0.45\textwidth}{!}{\includegraphics{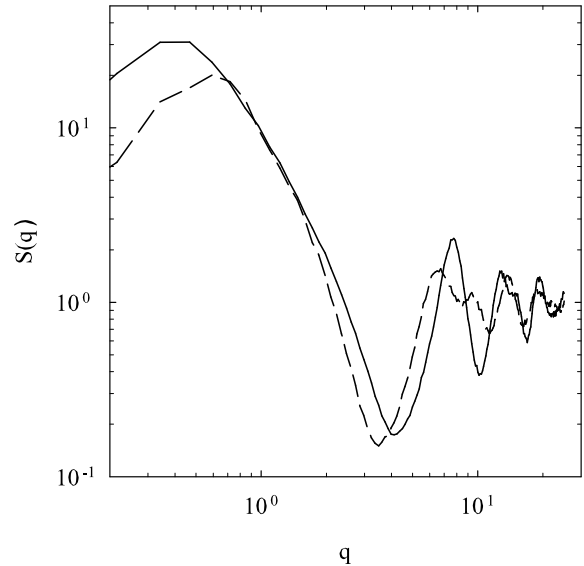}}
\caption{The $S(q)$ for $\epsilon=0.1$ (solid line) and $\epsilon=0.5$ (dashed line) at $\phi=0.05$.}
\label{f.12}
\end{figure}

For flexible DLCA the same effects are present, but in addition the system has a larger degree of freedom to increase the number of bonds by restructuring if $\epsilon$ is larger. As a consequence the value of $<z>$ of the gels increases with increasing $\epsilon$. For $\epsilon=0.5$ we found $<z> \approx 9.5$ after long times at low volume fractions. A qualitatively different local structure is formed when $\epsilon$ is larger than $0.41$ because the tetrahedral configuration is no longer the basic unit. Consequently all spheres are bound to all other spheres  for pentamers and hexamers. Figure(\ref{f.12}) shows a comparison of the structure factor at $\phi=0.05$ for $\epsilon=0.1$ and  $\epsilon=0.5$. The maximum is situated at larger $q$ for $\epsilon=0.5$ and has a smaller amplitude, indicating that the correlation length is smaller if $\epsilon$ is larger. The second maximum is situated at smaller $q$ indicating that the local structure (the strand thickness) is larger and the oscillations at high $q$ have a lower amplitude indicating less order. The comparison shows that when the interaction range is larger then the gels are more homogeneous with thicker strands and less local order. The effect of the interaction range on the structure is small for small $\epsilon$ and we expect that using a smaller interaction range than $0.1$ will have little influence on the structure.

\section{Discussion}
Experimentally, structures similar to those found in the present simulations of flexible DLCA have been reported for mixtures of hard spheres and polymers in which an effective attraction between the spheres is induced by depletion of the polymers \cite{779}. If the concentration of polymers is high the attraction may become so strong that the aggregation is irreversible on the time scale of observation. In this case no significant evolution of the structure is observed after gelation. We note that in experiments matters are sometimes complicated by electrostatic interaction. 

Recently, experiments on aggregating emulsion droplets were reported where the bonds were claimed to be truly irreversible, but bound droplets were free to move along the interface \cite{954}. The structure factor obtained from the experiments was compared to results of flexible DLA simulations \cite{seager}. Many of the features that distinguish the structure of clusters formed by rigid DLCA from those formed by flexible DLCA have also been observed in simulations of rigid and flexible DLA with zero interaction range (note that the expressions classic and slippery were used in \cite{seager} to refer to rigid and flexible, respectively). Also in DLA, flexible bonds caused the formation of thicker strands with a local tetrahedral configuration. As for DLCA, the fractal dimension of DLA clusters was the same for flexible and rigid bonds. The distribution of z was similar for flexible DLA and DLCA, and the structure factor showed prominent oscillations at high $q$-values in both cases. 

\begin{figure}[h]
\resizebox{0.45\textwidth}{!}{\includegraphics{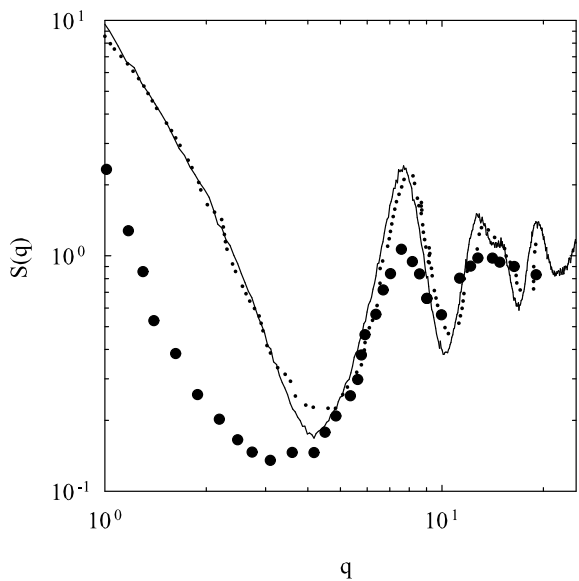}}
\caption{Comparison of $S(q)$ between flexible DLCA (solid line), slippery DLA \cite{seager}(dotted line) and experiment \cite{954} (circles).}
\label{f.13}
\end{figure}

In fig.(\ref{f.13}) the structure factor obtained from flexible DLCA ($\epsilon=0.1$) is compared to that obtained from flexible DLA ($\epsilon=0$) and experiments. The experimental structure factor was obtained by dividing the scattering data of the system with the particle form factor of the droplets. The latter was obtained experimentally before the aggregation had started and was assumed to be not influenced by interaction. As mentioned above, the aggregation process is more realistically described by DLCA, which results in a smaller fractal dimension ($d_f=1.8$) than obtained by DLA ($d_f=2.5$). Nevertheless, the results from the two types of simulations are remarkably close at large $q$-values, indicating that almost the same local structure is formed. This is perhaps not surprising if one considers that local restructuring occurred in both cases until the maximum number of bonds was formed under almost the same constraints. The simulations reproduced the position of the peak at $q=7.5$ seen in the experiments, but the amplitude was somewhat smaller (notice that the experimental data have been shifted upward for the comparison with the simulation results in figure 4 of \cite{seager}). The upturn at low $q$-values started at larger $q$ in the simulations than in the experiments. However, one should be careful when drawing conclusions on the basis of the deviation at low $q$-values, because the experimental results are very sensitive to the exact shape of the particle form factor in this $q$-range.

 As far as we are aware, flexible DLCA with strictly irreversible bonds has not been simulated before, but the present results may be compared with molecular dynamics simulations of reversible DLCA in the limit of very strong attraction. Foffi et al. \cite{883} studied the structure and dynamics of spheres with a square well interaction with width $\epsilon=0.005$ and two different well depths: $-2$ and $-20$ $kT$.  The interaction strengths may also be expressed in terms of the second virial coefficient \cite{892}: $B_2=-44$  and $B_2=-2.9\cdot 10^7$, in units of the particle volume. At both interaction strengths the systems are far below the binodal of the liquid-liquid phase separation. Crystallisation was avoided by using a bidisperse distribution of spheres with slightly different sizes. At smaller interaction strength the system slowly coarsened with time, but at $B2=-2.9\cdot 10^7$ the bonds may be considered irreversible on the time scale of the simulation. This situation is thus comparable to flexible DLCA.
  
The features of the gels obtained by Foffi et al. were very close to the ones reported here. Unfortunately, the authors did not show pair correlation functions, which would allow a more detailed comparison of the local structure. But the images of the clusters and the static structure factors of the gels resemble closely those shown here. The distribution of the bond coordination number was almost the same as the one found in this study, but $<z>$ was a bit smaller which can be explained by the narrower interaction range. Foffi et al. also observed a minimum of $<z>$ at volume fractions close to $25\%$, which they speculated to be related to the critical point of the liquid-liquid phase separation. We believe that the minimum is caused by a combination of opposing effects. At very low volume fractions the clusters can restructure while they grow, but with increasing volume fraction the growth rate increases and much of the restructuring has to occur after the percolation threshold. This increases the constraints on the restructuring and thus lowers $<z>$ with increasing $\phi$. At high volume fractions, however, the number of bonds increases due to crowding even for randomly distributed spheres. This leads to an increase of $<z>$ with increasing $\phi$. The latter effect is more important when $\epsilon$ is larger, which explains the stronger increase of $<z>$ found in the present study. 

Foffi et al. argued that the gel was formed by \textit{a phase separation process interrupted by attractive glass transition}. The same idea has also been put forward by others \cite{1030,manely,922,923,1029,1063,cardinaux}. It is clear, that gelation occurs only for interaction strengths where the equilibrium state would be phase separated and therefore one might indeed call the gel formation an interrupted phase separation. However, the suggested origin of the interruption is debatable. The concept of attractive glass was introduced to account for the slowing down of the dynamics at high concentrations with increasing attraction. Even if attractive glass formation would be a useful concept for hard spheres with a narrow interaction range at high volume fractions, it is not clear what it means in the context of the arrested state at low concentrations. Many particles are situated at the surface of the network strands and cannot be in a glassy state. These surface particles can escape from their neighbours by breaking all the bonds. This process inevitably leads to coarsening. The arrest is only inferred if the observation time is shorter than the time needed for surface particles to escape. We have done preliminary simulations that showed that the rate of coarsening decreased with decreasing $B_2$ following a power law and is simply too slow at $B_2=-2.9\cdot 10^7$ to observe on the time scale probed by Foffi et al.. 

The results presented here may also be compared with earlier Brownian dynamics simulations of colloidal gels formed by monodisperse spheres by d'Arjuzon et al.\cite{arju}. In those simulations the spheres interacted with a continuous potential that was steeply repulsive at contact and attractive over a range of $0.1$. The shape of the potential was chosen to be close to that of depletion interaction. Above a certain interaction strength the spheres crystallised \cite{463}, but when the interaction was very strong crystallisation was no longer observed for the duration of the simulation. A detailed study was done of the dynamics of this system at a single volume fraction $\phi=0.3$ and a minimum interaction potential of $-8kT$. For this interaction strength the life time of the bonds near the potential minimum is very long compared to the duration of the simulation. However, stretched bonds were still rapidly broken and reformed, because the potential goes to zero at the outer limit of the interaction range. In spite of the different shape of the potential and the reversibility of stretched bonds many features of this simulation are similar to those reported here for strictly irreversible aggregation with the same bond range. It was also found that $<z>$ increased rapidly at first followed by a very slow increase. A similar distribution of the bond coordination number was found at long times and $g(r)$ showed similar features notably the two peaks indicating tetrahedral structure. 
The MSD of spheres in the gel and also the distribution of displacements at  $\phi=0.3$ were similar to those reported here. The authors argued that displacements larger than one particle diameter were due to particles breaking of the network. It is clear from the present simulations, however, that even in irreversibly bound gels large scale mobility is possible due to cooperative motion. It is probable that at lower volume fractions the maximum displacement is determined by the correlation length of the gels. We note that in a recent simulation study of more open gels formed by restricting the bond angles and thus avoiding the formation of dense strands, it was also noted that the MSD displacement can be quite large due to cooperative motion \cite{delgado}. 

D'Arjuzon et al.  studied the self-intermediate scattering function in detail and found a fast decay towards a plateau followed by a slow decay to zero. The value of the plateau was shown to be directly related to $\delta$. The relaxation time of the slow decay increased with increasing waiting time and became very slow. Preliminary calculations showed similar behaviour for the square well system studied here. The similarities between the two systems indicate that the reversibility of stretched bonds is not an essential feature. We believe that in both systems the aging is caused by the formation of more (strong) bonds under the constraint that existing (strong) bonds cannot break at least for the duration of the simulation. 

	Lodge and Heyes \cite{420} studied spheres interacting with a Lennart-Jones potential with varying interaction range and depth using Brownian dynamics simulations. A similar shape of the pair correlation function was found if the range was narrow and the interaction strong. However, they also observed a peak at a smaller distance indicating that some crystallisation had occurred. The authors studied the phase separation kinetics up to $t=40$ in the time units used here. The kinetics was very slow in the case of strong attraction and may thus also be called aging. In fact the time dependence of the peak position shown in fig.(\ref{f.10}) is close to that obtained by Lodge and Heyes using a Lennart-Jones potential with a narrow interaction range. This indicates again that when the interaction is strong, the breaking of bonds is not important for the aging process at least in the early stages.  

\section{Conclusion}
Irreversible DLCA with flexible bonds causes locally densification, but on large length scales the structure is the same as for DLCA with rigid bonds. The fractal dimension of clusters formed in very dilute systems is $1.8$ in both cases. Locally the systems have a tetrahedral structure and show a certain degree of order that is independent of the volume fraction for $\phi<0.1$. At higher volume fraction the order is less distinct and similar to that of super cooled liquids. 

The system tries to maximize the number of bonds under the constraint of no bond breaking. The increase of the bond coordination number is fast at first, but it becomes progressively slower. The slow restructuring (aging) of the gels continues for very long times.  
At low volume fractions most of the local densification occurs while the clusters are formed, but at high volume fractions ($\phi>0.1$) the restructuring occurs mainly after gelation. 

At very low volume fractions the growth rate of the clusters during flexible DLCA is the same as for rigid DLCA and can be understood in terms of Smolechowski's kinetic equations. At high volume fractions the growth rate is somewhat faster, because flexibility increases the collision rate.

 	 Gels formed at low volume fractions show a large degree of flexibility. The average MSD stagnates at a value that increases with decreasing volume fraction. However, the MSD of the particles is highly heterogeneous reflecting the fractal structure of the gels. 
 	 
	Many features that appeared during irreversible flexible DLCA closely resemble simulations and experiments reported in the literature on spheres with a strong but finite attractive interaction. For the interpretation of these studies it is essential to consider whether the effect of bond breaking is significant during the time of observation.

\begin{acknowledgments}
This work has been supported in part by a grant from the Marie Curie Program
of the European Union numbered MRTN-CT-2003-504712.
\end{acknowledgments}

\end{document}